\documentclass[12pt, a4paper]{article}
\usepackage{amsmath,amssymb}
\usepackage{graphicx}
\usepackage{color}
\usepackage{cite,fancyhdr}
\usepackage[dvipdfmx]{hyperref}
\usepackage{subcaption}

\usepackage[affil-it]{authblk}
\usepackage[left=3cm,right=3cm,top=4cm,bottom=4cm,a4paper]{geometry}

\begin{document}
%
%
\setcounter{footnote}{0}
\setcounter{figure}{0}
\setcounter{table}{0}
\allowdisplaybreaks
\setcounter{footnote}{0}
\setcounter{figure}{0}
\setcounter{table}{0}


\title{\bf \large 
The sign of the dipole-dipole potential by axion exchange}
\author[1]{{\normalsize Ryuji Daido}}
\author[1,2]{{\normalsize Fuminobu Takahashi}}

\affil[1]{\small 
Department of Physics, Tohoku University,  

Sendai, Miyagi 980-8578, Japan}

\affil[2]{\small 
Kavli Institute for the Physics and Mathematics of the Universe (WPI),
 
University of Tokyo, Kashiwa 277--8583, Japan}

\date{}

\maketitle

\thispagestyle{fancy}
\rhead{TU-1041, IPMU17-0042 }
\cfoot{\thepage}
\renewcommand{\headrulewidth}{0pt}

\begin{abstract}
\noindent
We calculate a dipole-dipole potential  between fermions mediated by a light pseudoscalar, axion, 
paying a particular attention to the overall sign.  While the sign of the potential is  physical and important
for experiments to discover or constrain  the axion coupling to fermions,  there is often a sign error 
in the literature. The purpose of this short note is to clarify the sign issue of the axion-mediated dipole-dipole 
potential. As a by-product, we find a sign change of the dipole-dipole potenital due to the different spin
of the mediating particle.

\end{abstract}

\clearpage

\section{Introduction}
The exchange of a light particle gives rise to a force between other particles. 
One of such light particles is  a pseudo Nambu-Goldstone boson,  an axion, which
 appears in the Peccei-Quinn solution to the strong CP problem~\cite{Peccei:1977hh,Peccei:1977ur,Weinberg:1977ma,Wilczek:1977pj}.
The axion and axion-like particles have been searched for in many experiments
 (see e.g. Refs.~\cite{Kim:2008hd,Wantz:2009it,Ringwald:2012hr,Kawasaki:2013ae,Marsh:2015xka} 
 for recent reviews). 
 
 The axion or axion-like particle is generically coupled to nucleons and leptons, and the axion
exchange induces a spin-dependent force between them, which has been
constrained by many experiments~\cite{Olive:2016xmw}.
Furthermore,  there have been proposed many axion search experiments, 
 some of which aim to measure the spin-dependent force due to axion exchange 
 (e.g. \cite{Arvanitaki:2014dfa}). Therefore, it is important to unambiguously understand the relation between 
 the axion coupling and the observables.

In this short note we calculate a dipole-dipole potential  between fermions mediated by an axion, 
paying a particular attention to the overall sign.  The sign of the potential is physical and therefore
important for experiments to discover or constrain the axion coupling to fermions. In particular, 
the most challenging part of experiments searching for such axion-mediated dipole-dipole potential is 
to shield the standard magnetic dipole-dipole interaction to an extremely high degree. 
The experimental limits are therefore sensitive to the sign of the potential as well as the residual magnetic 
dipole-dipole interaction. 
We note however that there is often a sign error in the literature,
including the seminal paper~\cite{Moody:1984ba} on the axion-exchange potentials between fermions. 
Also, some of the experiments seem to use the potential with a wrong sign 
(e.g. \cite{Vasilakis:2008yn,Kotler:2015ura,Terrano:2015sna}).
The sign issue may not affect the experimental limits on the spin-dependent force
as long as the estimated limit is symmetric about zero. 
However, as a matter of fact, the sign is essential to claim a discovery of such a new force in future experiments.

The purpose of this short note is to clarify the sign issue of the dipole-dipole potential induced 
by axion exchange, 
and to eliminate the possibility that the sign error mediates to future experiments searching for such an 
anomalous force.

\section{Scalar exchange potential}
As an exercise, let us first consider a case where both fermions $\psi_1$ and $\psi_2$ interact with a real scalar $\phi$
through Yukawa interactions. The Lagrangian is given by
\begin{equation}
\mathcal{L}=\frac{1}{2}\partial_\mu\phi\partial^\mu\phi
-\frac{1}{2}m_\phi^2\phi^2 + \sum_{j=1,2} \left(\bar{\psi}_j(i\gamma^\mu\partial_\mu-M_j)\psi_j
- g_{Sj}\phi\,\bar{\psi}_j\psi_j \right)
\label{yukawa},
\end{equation} 
where $m_\phi$ and $M_j$ are the mass of $\phi$ and $\psi_j$, respectively, and $g_{Sj}$ is a Yukawa coupling
between $\phi$ and $\psi_j$.
Here and in what follows we adopt  the convention and notation used in the textbook by Peskin and 
Schroeder~\cite{Peskin:1995ev}, except for the representation of the gamma matrices and the normalization of the
spinors. We adopt the Dirac representation of the gamma matrices,
\begin{equation}
\gamma^0 = \left(
\begin{array}{cc}
1&0\\
0&-1
\end{array}
\right),~~
\gamma^i = \left(
\begin{array}{cc}
0&\sigma^i\\
-\sigma^i&0
\end{array}
\right),~{\rm and}~
\gamma_5 = \left(
\begin{array}{cc}
0&1\\
1&0
\end{array}
\right),~
\end{equation}
where $\sigma^i$ are the Pauli matrices.

\begin{figure}
\begin{minipage}{0.5\hsize}
\centering
\includegraphics[width=6cm]{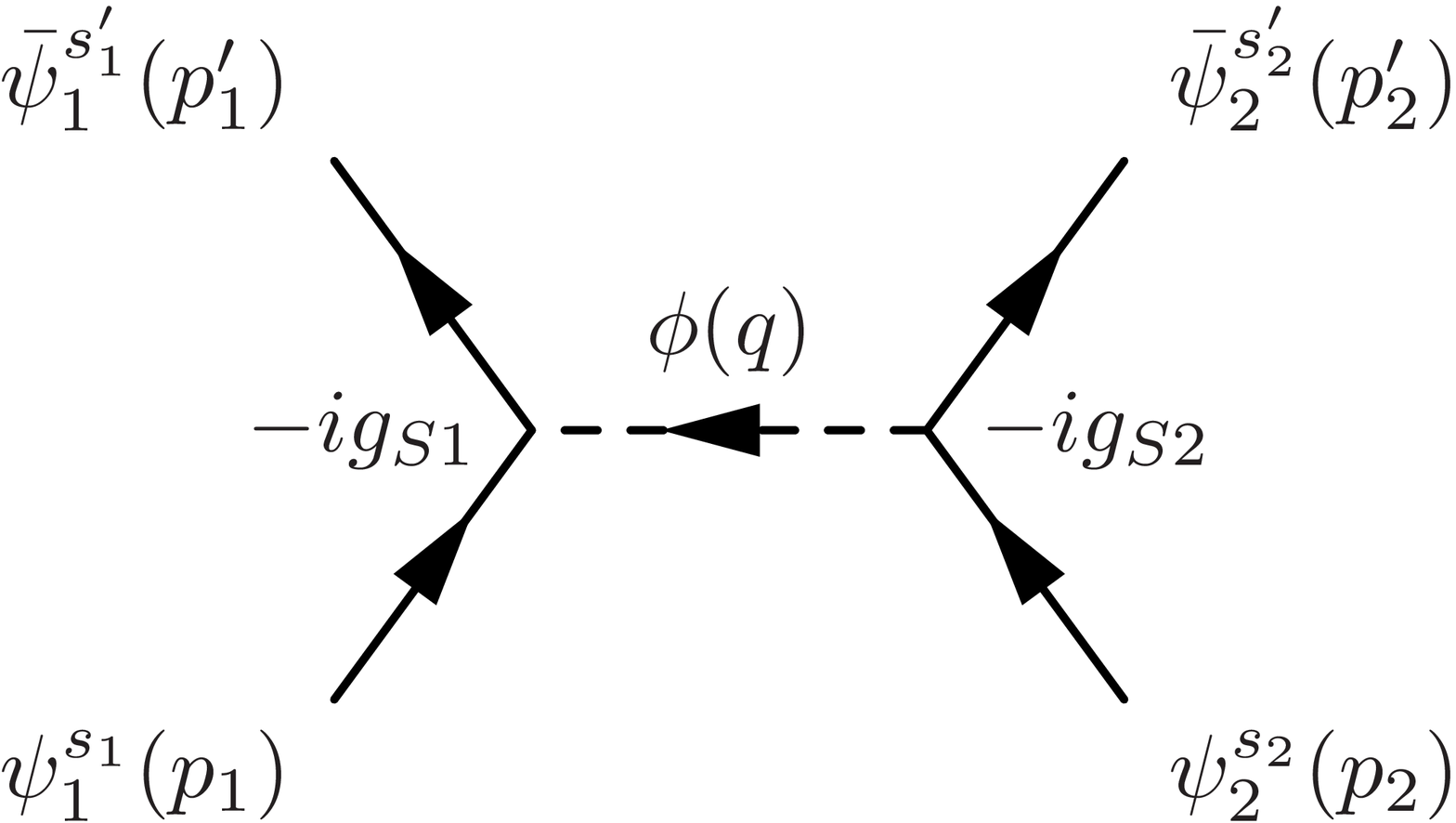}
\subcaption{Scalar exchange diagram}\label{diagram1}
\end{minipage}
\begin{minipage}{0.5\hsize}
\centering
\includegraphics[width=6cm]{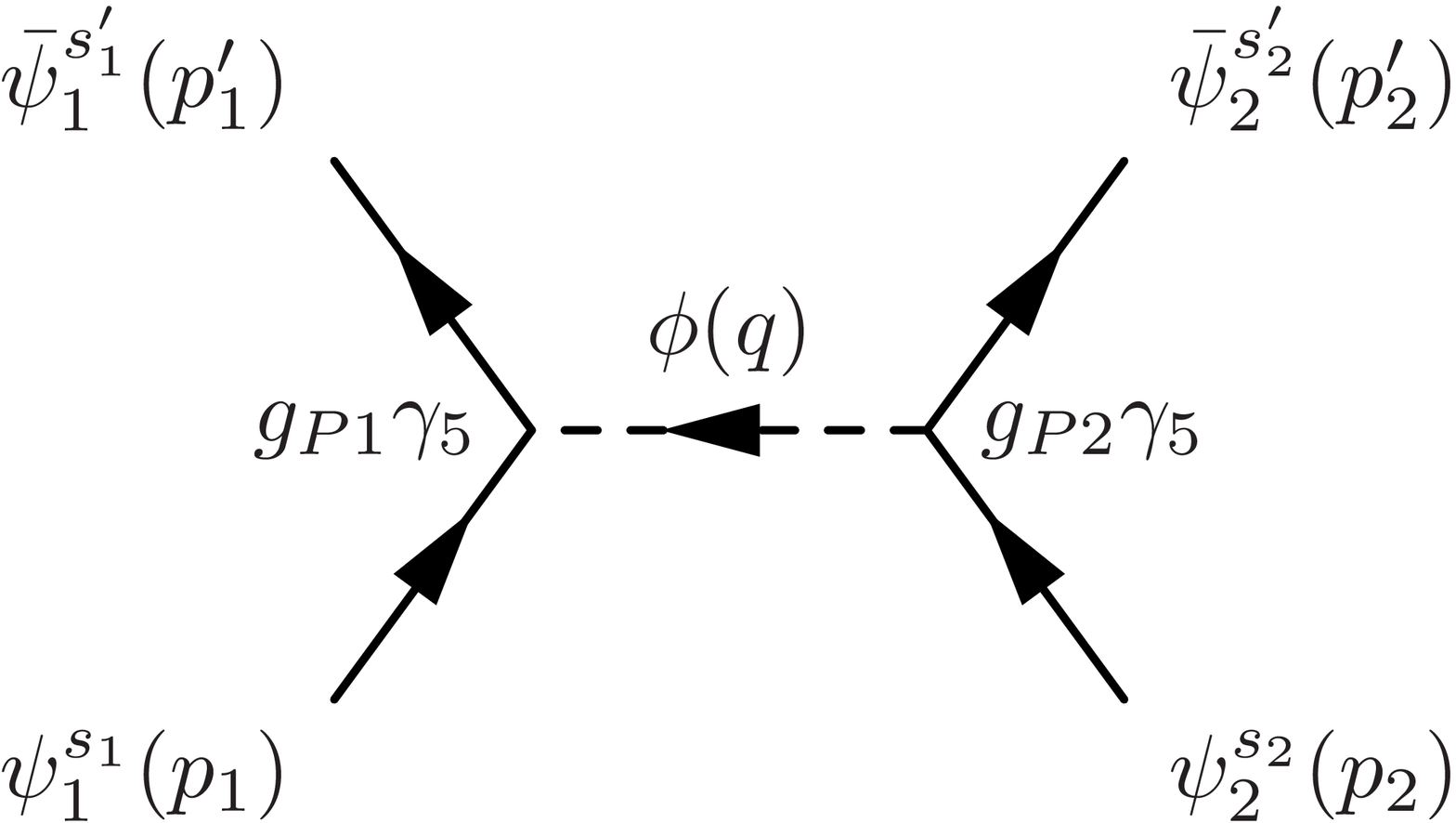}
\subcaption{Pseudoscalar exchange diagram}\label{diagram2}
\end{minipage}
\caption{Feynman diagrams for fermion scattering mediated by (a) scalar and (b) pseudo scalar, respectively.}
\end{figure}

The vertex factor of the Yukawa interaction is then given by
\begin{equation}
(-i )g_{Sj}.\label{v1}
\end{equation}
The scattering amplitude for $\psi_1(p_1)\psi_2(p_2)\to\psi_1(p_1')\psi_2(p_2')$ mediated 
by the scalar $\phi$ (see Fig. \ref{diagram1}) is expressed as
\begin{equation}
i\mathcal{M}=\bar{u}_1^{s_1'}(p_1')(-ig_{S1})u_1^{s_1}(p_1)\frac{i}{q^2-m_\phi^2}\bar{u}_2^{s_2'}(p_2')(-ig_{S2})u_2^{s_2}(p_2).\label{amp1}
\end{equation}
Here $q\equiv p_1'-p_1=-(p_2'-p_2)$ is the momentum transfer between 
the two fermions, and the superscripts $s_1$, $s_1'$, $s_2$ and $s_2'$ 
denote the spin of each fermion.  We adopt  the non-relativistic normalization 
condition ${u_j^{s'}}^\dag(p)u_j^s(p)=\delta_{s,s'}$ (no summation over $j$), which is different from
the usual relativistic normalization, ${u_j^{s'}}^\dag(p)u_j^s(p)= 2 E \,\delta_{s,s'}$, where $p = (E, \vec{p})$.
With the non-relativistic normalization, the plane wave solution is given by
\begin{equation}
u_j^s(p)=\sqrt{\frac{E+M_j}{2E}}\left(\chi_s,\, \frac{\vec{\sigma}\cdot\vec{p}}{E+M_j}\chi_s\right)^\mathrm{T}\simeq\left(\chi_s,\, \frac{\vec{\sigma}\cdot\vec{p}}{2M_j}\chi_s\right)^\mathrm{T},\label{plane}
\end{equation}
where $\chi_s$ is a  two component spinor satisfying $\chi^\dag_{s'}\chi_s=\delta_{s,s'}$, and
we have taken the non-relativistic limit, $E \approx M_j$,  in the second equality.
Substituting this solution into (\ref{amp1}), one obtains
\begin{equation}
i\mathcal{M} \simeq -i\frac{g_{S1}g_{S2}}{(q^0)^2-|\vec{q}|^2-m_\phi^2}\delta_{s_1,s_1'}\delta_{s_2,s_2'}\simeq i\frac{g_{S1}g_{S2}}{|\vec{q}|^2+m_\phi^2}\delta_{s_1,s_1'}\delta_{s_2,s_2'},
\end{equation}
where we have used, $|q^0| \ll |\vec{q}|$, in the non-relativistic limit in the second equality.

In general, the position-space potential can be obtained 
by taking the Fourier transform of the momentum-space amplitude with respect to 
the momentum transfer $\vec{q}$. In the case at hand, it is given by
\begin{align}
V(\vec{r})&=-\int\frac{d^3q}{(2\pi)^3} e^{i\vec{q}\cdot\vec{r}}
\left(\frac{g_{S1}g_{S2}}{|\vec{q}|^2+m_\phi^2}\right)
\\
&=-\frac{g_{S1}g_{S2}}{4\pi r} e^{-m_{\phi}r}
\label{yukawaV}
\end{align}
This result is of course consistent 
with Refs.~\cite{Moody:1984ba} and \cite{Dobrescu:2006au} and many other literatures. 

\section{Pseudoscalar exchange potential}
Next, let us go on to another type of interaction involving a pseudoscalar $\phi$. The Lagrangian is given by
\begin{equation}
\mathcal{L}=\frac{1}{2}\partial_\mu\phi\partial^\mu\phi-\frac{1}{2}m_\phi^2\phi^2
+\sum_{j=1,2} \left(
\bar{\psi}_j(i\gamma^\mu\partial_\mu-M_j)\psi_j-  i g_{Pj}\phi \,\bar{\psi}_j\gamma_5\psi_j
\right)
\label{pyukawa},
\end{equation} 
where $g_{Pj}$ is a real coupling constant between $\phi$ and $\psi_j$. Note that the imaginary unity $i$ is 
included in the last term to make $g_{Pj}$ real.  
The vertex factor of the above interaction is
\begin{equation}
(-i)ig_{Pj}\gamma_{5} = g_{Pj}\gamma_{5}.
\end{equation}
Then the scattering amplitude for $\psi_1(p_1)\psi_2(p_2)\to\psi_1(p_1')\psi_2(p_2')$ mediated by the pseudoscalar (see Fig. \ref{diagram2}) is given by
\begin{equation}
i\mathcal{M}=\bar{u}_1^{s_1'}(p_1')g_{P1}\gamma_5u_1^{s_1}(p_1)\frac{i}{q^2-m_\phi^2}\bar{u}_2^{s_2'}(p_2')g_{P2}\gamma_5u_2^{s_2}(p_2).\label{amp2}
\end{equation}
Using the plane wave solution (\ref{plane}), we can calculate the matrix element
\begin{align}
\bar{u}_1^{s_1'}(p_1')\gamma_5u_1^{s_1}(p_1) & \simeq \chi_{s_1'}^{\dag}\left(\frac{\vec{\sigma}}{2M_1}\cdot(\vec{p_1}-\vec{p_1}')\right)\chi_{s_1},\\
\bar{u}_2^{s_2'}(p_2')\gamma_5u_2^{s_2}(p_2) & \simeq \chi_{s_2'}^{\dag}\left(\frac{\vec{\sigma}}{2M_2}\cdot(\vec{p_2}-\vec{p_2}')\right)\chi_{s_2},
\end{align}
and the amplitude becomes
\begin{align}
i\mathcal{M}& \simeq i\frac{g_{P1}g_{P2}}{q^2-m_\phi^2}\frac{
\left(\chi^\dag_{s_1'}(\vec{\sigma}\cdot(-\vec{q}))\chi_{s_1}\right)
\left(\chi^{\dag}_{s_2'}(\vec{\sigma}\cdot\vec{q})\chi_{s_2} \right)
}{4M_1M_2},  \\
&\simeq i\frac{g_{P1}g_{P2}}{|\vec{q}|^2+m_\phi^2}\frac{
\left(\chi^\dag_{s_1'}(\vec{\sigma}\cdot\vec{q})\chi_{s_1} \right)
\left(\chi^{\dag}_{s_2'}(\vec{\sigma}\cdot\vec{q})\chi_{s_2}\right)
}{4M_1M_2},
\label{MP}
\end{align}
where we have taken the nonrelativistic limit in the second equality. 
Then, by taking the Fourier transform of the amplitude,  we obtain
\begin{align}
V(\vec{r})&=-\int\frac{d^3q}{(2\pi)^3} e^{i\vec{q}\cdot\vec{r}} \left[
\frac{g_{P1}g_{P2}}{|\vec{q}|^2+m_\phi^2}\frac{
(\vec{{ S}}_1\cdot\vec{q})(\vec{{ S}}_2\cdot\vec{q})
}{M_1M_2}\right],
\\
&=\frac{g_{P1}g_{P2}}{M_1M_2}(\vec{{ S}}_1\cdot\vec{\nabla})(\vec{{ S}}_2\cdot\vec{\nabla})\int\frac{d^3q}{(2\pi)^3}\frac{1}{|\vec{q}|^2+m_\phi^2} e^{i\vec{q}\cdot\vec{r}},\\
&=\frac{g_{P1}g_{P2}}{M_1M_2}(\vec{{ S}}_1\cdot\vec{\nabla})(\vec{{ S}}_2\cdot\vec{\nabla})\left(\frac{e^{-m_\phi r}}{4\pi r}\right),
\end{align}
where $\vec{{S}}_{1}$ and $\vec{{S}}_{2}$ are the spin operators of the fermions $\psi_1$ and $\psi_2$, 
respectively. In the literature, the spin operators are often represented by $\vec{\sigma}_1$ and $\vec{\sigma}_2$,
and they are related as $\vec{{S}}_{1} = \vec{\sigma}_1/2$ and $\vec{{S}}_{2} = \vec{\sigma}_2/2$.
Finally,  using the formula
\begin{equation}
\nabla_i\nabla_j\left(\frac{e^{-m_\phi r}}{r}\right)=-e^{-m_\phi r}\left[\delta_{ij}\left(\frac{m_\phi}{r^2}+\frac{1}{r^3}+\frac{4\pi}{3}\delta^3(r)\right)-\hat{r}_i\hat{r}_j\left(\frac{m_\phi^2}{r}+\frac{3m_\phi}{r^2}+\frac{3}{r^3}\right)\right],
\end{equation}
we arrive at the dipole-dipole potential induced by axion exchange,
\begin{align}
V(\vec{r})&=-\frac{g_{P1}g_{P2}\exp(-m_\phi r)}{4\pi M_1M_2}\left[(\vec{{S}}_{1}\cdot\vec{{S}}_{2})\left(\frac{m_\phi}{r^2}+\frac{1}{r^3}+\frac{4\pi}{3}\delta^3(r)\right) \right.\nonumber\\
&~~~~~~~~~~~~~~~~~~~~~~~~~~~~~~~~~~\left.-(\vec{{S}}_{1}\cdot\hat{r})(\vec{{S}}_{2}\cdot\hat{r})\left(\frac{m_\phi^2}{r}+\frac{3m_\phi}{r^2}+\frac{3}{r^3}\right)\right],
\label{vdp}
\end{align}
where $\hat{r}\equiv \vec{r}/r$ is the unit vector. 
In the massless limit, we obtain
\begin{align}
\label{massless}
V(\vec{r})&\to -\frac{g_{P1}g_{P2}}{4 \pi M_1M_2 r^3 }\left[
\vec{{S}}_{1}\cdot\vec{{S}}_{2}
- 3 (\vec{{S}}_{1}\cdot\hat{r})(\vec{{S}}_{2}\cdot\hat{r})
\right],~~~~~~(m_\phi \to 0)
\end{align}
where we have dropped the contact term. The form of the potential in the massless limit 
should be compared with the standard  magnetic dipole-dipole interaction,
\begin{align}
\label{magnetic}
H_{\mu \mu} = - \frac{3 (\vec{\mu}_1\cdot \hat{r}) (\vec{\mu}_2\cdot \hat{r}) 
- \vec{\mu}_1 \cdot \vec{\mu}_2}{4 \pi r^3},
\end{align}
where $\vec{\mu}_1$ and $\vec{\mu}_2$ are the magnetic moments and we have omitted
the Fermi contact term. The magnetic moment for an elementary Dirac particle is related to 
its spin as
\begin{align}
\vec{\mu} = g \frac{e}{2m} \vec{{S}},
\end{align}
where $m$ and $g$ are the mass and $g$-factor of the particle, respectively. 
Notice the overall sign difference between (\ref{magnetic}) and  (\ref{massless}) when $g_{P1} = g_{P2}$.\footnote{
It is well-known that the exchange of a scalar particle produces an attractive force (cf. Eq.~(\ref{yukawaV})),
of a spin $1$ particle (e.g. photon) a repulsive force between likes, and of a spin $2$ particle (graviton) an attractive force. 
We find that a similar sign change of the potential due to the different spin of the mediating particle arises
also for the spin-dependent force: the sign of the dipole-dipole potential mediated
by the graviton is same as (\ref{vdp})~\cite{Holstein:2008sx}.
We thank Georg Raffelt for pointing out 
this issue.
}

The above result (\ref{vdp}) (and (\ref{massless})) is
 consistent with Ref.~\cite{Dobrescu:2006au}, and with Ref.~\cite{Anselm:1986bz} in the limit of $m_\phi\to0$. 
It is also consistent with the one (neutral) pion exchange potential between nucleons~\cite{Brueckner:1953zzb}. 
On the other hand, the results in e.g. Refs.~\cite{Anselm:1982hw,Moody:1984ba} have an opposite sign. 

The sign of the potential is physical, therefore it is important for experiments to discover or constrain 
 the axion couplings
to fermions. Unfortunately,  there is often the sign error in the literature (including the equation in the header
 of the limit on invisible axion electron coupling in PDG~\cite{Olive:2016xmw}).
 So far, the axion-mediated dipole-dipole potential is only limited by experiments, and the sign error
 does not change the results unless the estimated error is antisymmetric about the sign of the potential. 
 However, as a matter of fact, the sign is essential to claim a discovery of such potential in future experiments.

\appendix
\section{Monopole-dipole potential by axion exchange}
Here we  give a monopole-dipole potential by the axion exchange for completeness. 
Assuming a scalar coupling to $\psi_1$ and a pseudoscalar coupling to $\psi_2$, we
obtain after a similar calculation, 
\begin{align}
V(\vec{r})&=-\int\frac{d^3q}{(2\pi)^3} e^{i\vec{q}\cdot\vec{r}} \left[
\frac{i g_{S1}g_{P2}}{|\vec{q}|^2+m_\phi^2}
\frac{(\vec{{S}}_2\cdot\vec{q})}{M_2}\right] 
\\
&=- \frac{g_{S1}g_{P2}}{M_2} (\vec{{S}}_2 \cdot \vec{\nabla})
\int\frac{d^3q}{(2\pi)^3}\frac{1}{|\vec{q}|^2+m_\phi^2} e^{i\vec{q}\cdot\vec{r}}\\
&=\frac{g_{S1}g_{P2}}{4 \pi M_2} (\vec{{S}}_2 \cdot \hat{r}) \left(\frac{m_\phi}{r}+\frac{1}{r^2}\right)
e^{-m_\phi r},
\end{align}
which agrees with the result in Ref.~\cite{Moody:1984ba}.

\section*{Acknowledgments}
We are grateful to G. Raffelt and A. Ringwald for reading the manuscript and providing useful comments.
We also thank Y. Fujitani and N. Yokozaki for checking the calculations. 
This work is supported by Tohoku University Division for Interdisciplinary Advanced Research and Education (R.D.),  JSPS KAKENHI Grant Numbers JP15H05889 (F.T.), JP15K21733 (F.T.), 
JP26247042 (F.T),  JP26287039 (F.T.) and by World Premier International Research 
Center Initiative (WPI Initiative), MEXT, Japan (F.T.).

\bibliographystyle{utphys}
\bibliography{reference}

\providecommand{\href}[2]{#2}\begingroup\raggedright\begin{thebibliography}{10}

\bibitem{Peccei:1977hh}
R.~D. Peccei and H.~R. Quinn, ``{CP Conservation in the Presence of
  Instantons},''
\href{http://dx.doi.org/10.1103/PhysRevLett.38.1440}{{\em Phys. Rev. Lett.}
  {\bfseries 38} (1977) 1440--1443}.

\bibitem{Peccei:1977ur}
R.~D. Peccei and H.~R. Quinn, ``{Constraints Imposed by CP Conservation in the
  Presence of Instantons},''
\href{http://dx.doi.org/10.1103/PhysRevD.16.1791}{{\em Phys. Rev.} {\bfseries
  D16} (1977) 1791--1797}.

\bibitem{Weinberg:1977ma}
S.~Weinberg, ``{A New Light Boson?},''
\href{http://dx.doi.org/10.1103/PhysRevLett.40.223}{{\em Phys. Rev. Lett.}
  {\bfseries 40} (1978) 223--226}.

\bibitem{Wilczek:1977pj}
F.~Wilczek, ``{Problem of Strong p and t Invariance in the Presence of
  Instantons},''
\href{http://dx.doi.org/10.1103/PhysRevLett.40.279}{{\em Phys. Rev. Lett.}
  {\bfseries 40} (1978) 279--282}.

\bibitem{Kim:2008hd}
J.~E. Kim and G.~Carosi, ``{Axions and the Strong CP Problem},''
  \href{http://dx.doi.org/10.1103/RevModPhys.82.557}{{\em Rev. Mod. Phys.}
  {\bfseries 82} (2010) 557--602},
\href{http://arxiv.org/abs/0807.3125}{{\ttfamily arXiv:0807.3125 [hep-ph]}}.

\bibitem{Wantz:2009it}
O.~Wantz and E.~P.~S. Shellard, ``{Axion Cosmology Revisited},''
  \href{http://dx.doi.org/10.1103/PhysRevD.82.123508}{{\em Phys. Rev.}
  {\bfseries D82} (2010) 123508},
\href{http://arxiv.org/abs/0910.1066}{{\ttfamily arXiv:0910.1066
  [astro-ph.CO]}}.

\bibitem{Ringwald:2012hr}
A.~Ringwald, ``{Exploring the Role of Axions and Other WISPs in the Dark
  Universe},'' \href{http://dx.doi.org/10.1016/j.dark.2012.10.008}{{\em Phys.
  Dark Univ.} {\bfseries 1} (2012) 116--135},
\href{http://arxiv.org/abs/1210.5081}{{\ttfamily arXiv:1210.5081 [hep-ph]}}.

\bibitem{Kawasaki:2013ae}
M.~Kawasaki and K.~Nakayama, ``{Axions: Theory and Cosmological Role},''
  \href{http://dx.doi.org/10.1146/annurev-nucl-102212-170536}{{\em Ann. Rev.
  Nucl. Part. Sci.} {\bfseries 63} (2013) 69--95},
\href{http://arxiv.org/abs/1301.1123}{{\ttfamily arXiv:1301.1123 [hep-ph]}}.

\bibitem{Marsh:2015xka}
D.~J.~E. Marsh, ``{Axion Cosmology},''
  \href{http://dx.doi.org/10.1016/j.physrep.2016.06.005}{{\em Phys. Rept.}
  {\bfseries 643} (2016) 1--79},
\href{http://arxiv.org/abs/1510.07633}{{\ttfamily arXiv:1510.07633
  [astro-ph.CO]}}.

\bibitem{Olive:2016xmw}
{\bfseries Particle Data Group} Collaboration, C.~Patrignani {\em et~al.},
  ``{Review of Particle Physics},''
\href{http://dx.doi.org/10.1088/1674-1137/40/10/100001}{{\em Chin. Phys.}
  {\bfseries C40} no.~10, (2016) 100001}.

\bibitem{Arvanitaki:2014dfa}
A.~Arvanitaki and A.~A. Geraci, ``{Resonantly Detecting Axion-Mediated Forces
  with Nuclear Magnetic Resonance},''
  \href{http://dx.doi.org/10.1103/PhysRevLett.113.161801}{{\em Phys. Rev.
  Lett.} {\bfseries 113} (2014) 161801},
\href{http://arxiv.org/abs/1403.1290}{{\ttfamily arXiv:1403.1290 [hep-ph]}}.

\bibitem{Moody:1984ba}
J.~E. Moody and F.~Wilczek, ``{New Macroscopic Forces?},''
\href{http://dx.doi.org/10.1103/PhysRevD.30.130}{{\em Phys. Rev.} {\bfseries
  D30} (1984) 130}.

\bibitem{Vasilakis:2008yn}
G.~Vasilakis, J.~M. Brown, T.~W. Kornack, and M.~V. Romalis, ``{Limits on new
  long range nuclear spin-dependent forces set with a K - He-3
  co-magnetometer},''
  \href{http://dx.doi.org/10.1103/PhysRevLett.103.261801}{{\em Phys. Rev.
  Lett.} {\bfseries 103} (2009) 261801},
\href{http://arxiv.org/abs/0809.4700}{{\ttfamily arXiv:0809.4700
  [physics.atom-ph]}}.

\bibitem{Kotler:2015ura}
S.~Kotler, R.~Ozeri, and D.~F.~J. Kimball, ``{Constraints on exotic
  dipole-dipole couplings between electrons at the micrometer scale},''
  \href{http://dx.doi.org/10.1103/PhysRevLett.115.081801}{{\em Phys. Rev.
  Lett.} {\bfseries 115} no.~8, (2015) 081801},
\href{http://arxiv.org/abs/1501.07891}{{\ttfamily arXiv:1501.07891
  [physics.atom-ph]}}.

\bibitem{Terrano:2015sna}
W.~A. Terrano, E.~G. Adelberger, J.~G. Lee, and B.~R. Heckel, ``{Short-range
  spin-dependent interactions of electrons: a probe for exotic pseudo-Goldstone
  bosons},'' \href{http://dx.doi.org/10.1103/PhysRevLett.115.201801}{{\em Phys.
  Rev. Lett.} {\bfseries 115} no.~20, (2015) 201801},
\href{http://arxiv.org/abs/1508.02463}{{\ttfamily arXiv:1508.02463 [hep-ex]}}.

\bibitem{Peskin:1995ev}
M.~E. Peskin and D.~V. Schroeder, {\em {An Introduction to quantum field
  theory}}.
\newblock 1995.
\newblock
\url{http://www.slac.stanford.edu/spires/find/books/www?cl=QC174.45%3AP4}.
\newblock

\bibitem{Dobrescu:2006au}
B.~A. Dobrescu and I.~Mocioiu, ``{Spin-dependent macroscopic forces from new
  particle exchange},''
  \href{http://dx.doi.org/10.1088/1126-6708/2006/11/005}{{\em JHEP} {\bfseries
  11} (2006) 005},
\href{http://arxiv.org/abs/hep-ph/0605342}{{\ttfamily arXiv:hep-ph/0605342
  [hep-ph]}}.

\bibitem{Holstein:2008sx}
B.~R. Holstein and A.~Ross, ``{Spin Effects in Long Range Gravitational
  Scattering},''
\href{http://arxiv.org/abs/0802.0716}{{\ttfamily arXiv:0802.0716 [hep-ph]}}.

\bibitem{Anselm:1986bz}
A.~A. Anselm,
``{Massless Higgs-Goldstone bosons},''.

\bibitem{Brueckner:1953zzb}
K.~A. Brueckner and K.~M. Watson, ``{Nuclear Forces in Pseudoscalar Meson
  Theory},''
\href{http://dx.doi.org/10.1103/PhysRev.92.1023}{{\em Phys. Rev.} {\bfseries
  92} (1953) 1023--1035}.

\bibitem{Anselm:1982hw}
A.~A. Anselm, ``{Possible new long range interaction and methods for its
  observation},''
{\em JETP Lett.} {\bfseries 36} (1982) 55.

\end{thebibliography}\endgroup
\end{document}